\documentstyle[11pt]{article}
\begin{document}

\begin{titlepage}

\begin{flushright}
UQMATH-94-07\\
cond-mat/9410026\\
{\tt (Revised Version)}
\end{flushright}

\begin{center}
{\huge A New Supersymmetric and Exactly Solvable Model of Correlated Electrons}
\vskip.3in
{\Large Anthony J. Bracken, Mark D. Gould, Jon R. Links}\\
 and\\ {\Large Yao-Zhong Zhang}
 \footnote{Address after March 16, 1995:
 Yukawa Institute of Theoretical Physics, Kyoto University, Kyoto, Japan.}
\vskip.2in
{\large Department of Mathematics, University of Queensland, Brisbane,
Qld 4072,  Australia}
email: \{jrl,~yzz\}@maths.uq.oz.au
\end{center}
\vskip 2cm
\begin{center}
{\bf Abstract:}
\end{center}
A new lattice model is presented for correlated electrons
on the unrestricted $4^L$-dimensional electronic Hilbert space
$\otimes_{n=1}^L{\bf C}^4$ (where $L$ is the lattice length). It is
a supersymmetric generalization of the Hubbard model, but differs from
the extended Hubbard model proposed by Essler, Korepin and Schoutens.
The supersymmetry algebra of the new model is superalgebra $gl(2|1)$.
The model contains one symmetry-preserving free real parameter which is
the Hubbard interaction parameter $U$, and has its origin here in the
one-parameter family of inequivalent typical 4-dimensional irreps of $gl(2|1)$.
On a one-dimensional lattice, the model is exactly solvable by
the Bethe ansatz.


\end{titlepage}


\def\a{\alpha}
\def\b{\beta}
\def\d{\delta}
\def\e{\epsilon}
\def\g{\gamma}
\def\k{\kappa}
\def\l{\lambda}
\def\o{\omega}
\def\t{\theta}
\def\s{\sigma}
\def\D{\Delta}
\def\L{\Lambda}


\def\beq{\begin{equation}}
\def\eeq{\end{equation}}
\def\bea{\begin{eqnarray}}
\def\eea{\end{eqnarray}}
\def\ba{\begin{array}}
\def\ea{\end{array}}
\def\no{\nonumber}
\def\le{\langle}
\def\re{\rangle}

The Hubbard model and the $t$-$J$ model, both models of
correlated electrons on a lattice, and exactly solvable in one dimension,
have been extensively studied due to
their promising role in theoretical condensed-matter physics and
possibly in high-$T_c$ superconductivity.
The $t$-$J$ model is a lattice model on
the restricted $3^L$-dimensional electronic Hilbert space
$\otimes^L_{n=1}{\bf C}^3$ (throughout the paper, $L$ is the
lattice length), where the
occurrence of two electrons on the same lattice site is forbidden.
With the special choice of parameters: $t=1$ and $J=2$, the $t$-$J$
model becomes supersymmetric with the symmetry algebra being the superalgebra
$gl(2|1)$ \cite{Ess92b,Sar91}. In \cite{Ess92a,book,Ess??}, Essler,
Korepin and Schoutens (EKS) proposed a model, the so-called extended
Hubbard model, of correlated electrons on the unrestricted
$4^L$-dimensional electronic Hilbert space $\otimes_{n=1}^L{\bf C}^4$
. This EKS model, which allows doubly occupied sites and
combines and extends some of the interesting
features of the Hubbard model and the $t$-$J$ model,
is exactly solvable in one
dimension and has $gl(2|2)$ supersymmetry.

In this Letter, we propose another direction of generalization of the
Hubbard model.
Specifically, we propose a new model
on the same unrestricted $4^L$-dimensional electronic Hilbert space
$\otimes_{n=1}^L{\bf C}^4$, but with quite different interaction terms
from the ones in the EKS model.
Our model has $gl(2|1)$ supersymmetry and contains one symmetry-preserving
free real parameter which is exactly the Hubbard
interaction parameter $U$;
this real parameter $U$ has its origin here in
the one-parameter family of inequivalent typical 4-dimensional irreps of
$gl(2|1)$. The model can naturally be regarded as a modified
Hubbard model with additional nearest-neighbor interactions and
is again exactly solvable on a one dimensional lattice. The exact
solvability of our model in one dimension comes from the fact
that as an abstract
dynamical model it is derived from a $gl(2|1)$-invariant rational R-matrix
which satisfies the (graded) quantum Yang-Baxter equation (QYBE).

It seems that only a $gl(2|1)$-symmetric lattice model on
the unrestricted $4^L$-dimensional electronic Hilbert space
could be a natural candidate for
the lattice analogue of $N$=2 superconformal field theory, of which
the $gl(2|1)=osp(2|2)$ algebra defines the underlying symmetry, and which
is a critically fixed point of the $N$=2 supersymmetric Landau-Ginzburg
model \cite{Gomez}. This gives another motivation for our model.

Let us begin by introducing some notation as in \cite{Ess92a}.
Electrons on a lattice are described by canonical Fermi operators $c_{i,\s}$
and $c_{i,\s}^\dagger$ satisfying the anti-commutation relations given by
$\{c_{i,\s}^\dagger, c_{j,\tau}\}=\d_{ij}\d_{\s\tau}$, where
$i,j=1,2,\cdots,L$ and $\s,\tau=\uparrow,\;\downarrow$.
The operator $c_{i,\s}$ annihilates an electron of spin $\s$
at site $i$, which implies that the Fock vacuum $|0\re$ satisfies
$c_{i,\s}|0>=0$.
At a given lattice site $i$ there are four possible electronic states:
\beq
|0\re\,,~~~|\uparrow\re_i=c_{i,\uparrow}^\dagger|0\re\,,~~~
  |\downarrow\re_i=c_{i,\downarrow}^\dagger|0\re\,,~~~
  |\uparrow\downarrow\re_i=c_{i,\downarrow}^\dagger
  c_{i,\uparrow}^\dagger|0\re\,.\label{states}
\eeq
By $n_{i,\s}=c_{i,\s}^\dagger c_{i,\s}$ we denote the number operator for
electrons with spin $\s$ on site $i$, and we write $n_i=n_{i,\uparrow}+
n_{i,\downarrow}$. The spin operators $S\,,~S^\dagger\,,~
S^z$, (in the following, the global operator ${\cal O}$ will be always
expressed in terms of the local one ${\cal O}_i$ as
${\cal O}=\sum_{i=1}^L{\cal O}_i$ in one dimension)
\beq
S_i=c_{i,\uparrow}^\dagger c_{i,\downarrow}\,,~~~
  S_i^\dagger=c_{i,\downarrow}^\dagger c_{i,\uparrow}\,,~~~
  S_i^z=\frac{1}{2}(n_{i,\downarrow}-n_{i,\uparrow})\,,\label{s-operator}
\eeq
form an $sl(2)$ algebra and they commute with the hamiltonians that we
consider below.

In what follows, we only consider periodic lattice of length $L$.
The well-known Hubbard model hamiltonian takes the following form:
\bea
H^{\rm Hubbard}(U)
  &=&-\sum_{<i,j>}\;\sum_{\s=\uparrow,\downarrow}
  (c_{i,\s}^\dagger c_{j,\s}
 +c_{j,\s}^\dagger c_{i,\s})\no\\
& & +U\sum_{<i,j>}\left [ (n_{i,\uparrow}-\frac{1}{2})
 (n_{i,\downarrow}-\frac{1}{2})
 +(n_{j,\uparrow}-\frac{1}{2})
 (n_{j,\downarrow}-\frac{1}{2})\right ]\label{h-hubbard}
\eea
where $<i,j>$ denote nearest neighour links on the lattice. It contains
the hopping term for electrons and an on-site interaction term for
electron pairs (coupling $U$).

In \cite{Ess92a}, Essler et al
proposed a supersymmetric generalization of the Hubbard model.
The supersymmetry algebra in their model is $gl(2|2)$.
We present here another supersymmetric generalization of the Hubbard model.
The hamiltonian for our new model on a general $d$-dimensional
lattice reads
\bea
H^{\rm Q}(U)&\equiv& \sum_{<i,j>}H^{\rm Q}_{i,j}(U)=H^{\rm Hubbard}(U)
 +\frac{U}{2}\sum_{<i,j>}\sum_{\s=\uparrow,\downarrow}
 (c_{i,\s}^\dagger c_{i,-\s}
 ^\dagger c_{j,-\s} c_{j,\s}+{\rm h.c.})\no\\
& &+(1+\frac{U}{|U|}\sqrt{U+1})\sum_{<i,j>}\sum_{\s=\uparrow,\downarrow}
 (c_{i,\s}^\dagger c_{j,\s}
 +c_{j,\s}^\dagger c_{i,\s})(n_{i,-\s}+n_{j,-\s})\no\\
& &-\left (1+\frac{U}{|U|}\sqrt{U+1}\right )^2
 \sum_{<i,j>}\sum_{\s=\uparrow,\downarrow}
 (c_{i,\s}^\dagger c_{j,\s}+
 c_{j,\s}^\dagger c_{i,\s})n_{i,-\s}n_{j,-\s}\no\\
& &+\frac{U+2}{2}\sum_{<i,j>}(n_i+n_j).\label{brisbane}
\eea
As will be seen, the supersymmetry algebra underlying this model is
$gl(2|1)$. Remarkably, the model still contains the parameter
$U$ as a free parameter without breaking the supersymmetry. Also this
model is exactly solvable on the one dimensional periodic lattice, as
is seen below. Throughout this paper, we will restrict $U$ to the
range $U>-1$.

The hamiltonian (\ref{brisbane}) is obviously invariant under spin-reflection
$c_{i,\uparrow}\leftrightarrow c_{i,\downarrow}$.
It can be viewed as an extended Hubbard model with additional
nearest-neighbor interaction terms
in a different fashion from the one proposed in
\cite{Ess92a}. The physical nature of the additional terms is the following.
The second term is nothing but a pair-hopping term. The third and fourth
terms are bond-charge two-body and bond-charge-charge three-body
interaction terms, respectively. And the last term is
just a chemical potential. Clearly one can add to the above hamiltonian an
arbitrary chemical potential (coefficient $\mu$) term $\mu\sum_{i}n_i$ and an
external magnetic field (coefficient $h$) term
$h\sum_i(n_{i,\downarrow}-n_{i,\uparrow})$,
which commute with $H^{\rm Q}(U)$ but break its $gl(2|1)$ supersymmetry.

An interesting feature of our model is the discontinuity at $U=0$
. When $U\rightarrow 0^+$, the hamiltonian (\ref{brisbane})
reduces to
\bea
H^{\rm Q}(0^+)&=&-\sum_{<i,j>}\;\sum_{\s=\uparrow,\downarrow}
  (c_{i,\s}^\dagger c_{j,\s}
 +c_{j,\s}^\dagger c_{i,\s})\no\\
& &+2\sum_{<i,j>}\sum_{\s=\uparrow,\downarrow}
 (c_{i,\s}^\dagger c_{j,\s}
 +c_{j,\s}^\dagger c_{i,\s})(n_{i,-\s}-n_{j,-\s})^2\no\\
& &+\sum_{<i,j>}(n_i+n_j),\label{u=0}
\eea
containing a hopping term plus a bond-charge interaction term (up to a
chemical potential). Whereas as $U\rightarrow 0^-$, only a hopping term
(and a chemical potential) survives in the hamiltonian (\ref{brisbane}).

Our local hamiltonian $H^{\rm Q}_{i,j}(U)$ does not act as graded
permutation of the electron states (\ref{states}) at sites $i$ and $j$,
in contrast to the hamiltonian in \cite{Ess92a}.
Nevertheless, it is supersymmetric, and the global hamiltonian commutes
with global number operators of spin up and spin down, respectively.
There are four supersymmetry generators for $H^{\rm Q}(U)$:
$Q_\uparrow,~Q^\dagger_\uparrow,~
Q_\downarrow$, and $Q_\downarrow^\dagger$ with the
corresponding local operators given by
\bea
&&Q_{i,\uparrow}=-\sqrt{\a}\,n_{i,\downarrow}c_{i,\uparrow}+\sqrt{\a+1}\,
  (1-n_{i,\downarrow})c_{i,\uparrow},\no\\
&&Q_{i,\downarrow}=-\sqrt{\a}\,
  n_{i,\uparrow}c_{i,\downarrow}+\sqrt{\a+1}
  \,(1-n_{i,\uparrow})c_{i,\downarrow}\label{super-Q}
\eea
where $0\leq {\rm arg}\sqrt{Z}<\pi,~Z=\a$ or $\a+1$,
and $\a\geq 0$ or $\a<-1$ is the inverse of $U$:
\beq
\a=\frac{1}{U}.
\eeq
These generators, together with $S,~S^\dagger,~S^z$ and two others
($E^2_2+E^3_3$ and $E^3_3$, defined below), form the superalgebra $gl(2|1)$.
To prove this, we denote the generators of $gl(2|1)$ by $E^\b_\g,~~
\b,\g=1,2,3$ with grading $[1]=[2]=0,~[3]=1$. In a typical 4-dimensional
representation of $gl(2|1)$, the highest weight itself of the
representation depends on the free parameter $\a$, thus giving rise to
a one-parameter family of inequivalent irreps \cite{Kac}.
Choose the following basis
\beq
|4\re=\left (
\begin{array}{c}
0\\
0\\
0\\
1
\end{array} \right ),~~~
|3\re=\left (
\begin{array}{c}
0\\
0\\
1\\
0
\end{array} \right ),~~~
|2\re=\left (
\begin{array}{c}
0\\
1\\
0\\
0
\end{array} \right ),~~~
|1\re=\left (
\begin{array}{c}
1\\
0\\
0\\
0
\end{array} \right )
\eeq
with $|1\re,~|4\re$ even (bosonic) and $|2\re,~|3\re$ odd (fermionic).
Then in this typical 4-dimensional representation,
$E^\b_\g$ are $4\times 4$ supermatrices of the form
\bea
&&E^1_2=|2\re \le 3|,~~~E^2_1=|3\re\le 2|,~~~E^1_1=-|3\re\le 3|-|4\re\le 4|,~~~
  E^2_2=-|2\re\le 2|-|4\re\le 4|,\no\\
&&E^2_3=\sqrt{\a}\,|1\re\le 2|+\sqrt{\a+1}\,|3\re\le 4|,~~~
  E^3_2=\sqrt{\a}\,|2\re\le 1|+\sqrt{\a+1}\,|4\re\le 3|,\no\\
&&E^1_3=-\sqrt{\a}\,|1\re\le 3|+\sqrt{\a+1}\,|2\re\le 4|,~~~
  E^3_1=-\sqrt{\a}\,|3\re\le 1|+\sqrt{\a+1}\,|4\re\le 2|,\no\\
&&E^3_3=\a\,|1\re\le 1|+(\a+1)\,\left (|2\re\le 2|+|3\re\le 3|\right )
  +(\a+2)\,|4\re\le 4|.\label{matrix}
\eea
For $\a>0$,
\beq
\left (E^\b_\g\right )^\dagger=E^\g_\b
\eeq
and we call the representation unitary of type I. For $\a<-1$, we
have
\beq
\left (E^\b_\g\right )^\dagger=(-1)^{[\b]+[\g]}\;E^\g_\b
\eeq
and we refer to the representation as unitary of type II. In this paper,
we are interested in these unitary representations. For a description
and classification of the two types of unitary representations, see
\cite{Mar90}.

Further choosing
\beq
|4\re\equiv |0\re,~~~~
|3\re\equiv |\uparrow\re,~~~~
|2\re\equiv |\downarrow\re,~~~~
|1\re\equiv |\uparrow\downarrow\re\label{choice}
\eeq
and with the help of the following identities,
\bea
&&|0\re\le 0|+|\downarrow\re\le\downarrow|+|\uparrow\re\le\uparrow|
  +|\uparrow\downarrow\re\le\uparrow\downarrow|=1,\no\\
&&|\uparrow\downarrow\re\le\uparrow\downarrow|=n_\uparrow n_\downarrow,\no\\
&&|\uparrow\re\le\uparrow|=n_\uparrow-n_\uparrow n_\downarrow,\no\\
&&|\downarrow\re\le\downarrow|=n_\downarrow-n_\uparrow n_\downarrow,
\eea
one can easily establish that
\bea
&&n_i=(\a+2)-\left(E^3_3\right)_i,~~~~n_{i,\downarrow}=\left(E^1_1\right)_i
  +1,~~~~n_{i,\uparrow}=\left(E^2_2\right)_i+1,\label{nnn}\\
&&S_i^\dagger=\left (E^1_2\right )_i,~~~~S_i=\left (E_1^2\right )_i,~~~~
  S_i^z=\left (E^1_1\right )_i-\left (E^2_2\right )_i,\no\\
&&Q_{i,\downarrow}^\dagger=\left (E^3_2\right )^\dagger_i,~~~~
  Q_{i,\downarrow}=\left (E^3_2\right )_i,~~~~
  Q_{i,\uparrow}^\dagger=\left (E^3_1\right )^\dagger_i,~~~~
  Q_{i,\uparrow}=\left (E^3_1\right )_i.
\eea
The verification that the hamiltonian $H^{\rm Q}(U)$ commutes with all
nine generators of $gl(2|1)$ is just a straightforward calculation.
Eq.(\ref{nnn}) makes it clear that $H^{\rm Q}(U)$ commutes with the
global number operators of spin up and spin down, respectively.

The model is exactly solvable in one
dimension by the Bethe ansatz. To show this, we first of all show that
the local hamiltonian $H_{i,i+1}^{\rm Q}(U)$ on the one
dimensional lattice is actually derived from a $gl(2|1)$-invariant
rational R-matrix which satisfies the (graded) QYBE.
To this end, let $U_q[gl(2|1)]$  be the
well-known quantum (or $q$) deformation of $gl(2|1)$ and
$V$ be the $U_q[gl(2|1)]$-module with highest weight $(0,0|\a)$, which
affords the $q$-deformed version of the one-parameter family
of the inequivalent
typical 4-dimensional irreps \cite{Bra94a,Del94}. Without loss of
generality, we assume $q$ to be real. We also assume $q$ to be generic, i.e.
it is not a root of unity.
For $\a> 0$
or $\a<-1$, the module $V$ is unitary of type I and of type II,
respectively, and thus the tensor product $V\otimes V$
is completely reducible. We
write $V\otimes V=V_1\bigoplus V_2\bigoplus V_3$, where $V_1,~V_2$
and $V_3$ are $U_q[gl(2|1)]$-modules with  highest weights
$(0,0|2\a),~(0,-1|2\a+1)$ and $(-1,-1|2\a+2)$, respectively \cite{Bra94a},
and let $\check{P}_k,~k=1,2
,3$ be the projection operator from $V\otimes V$ onto $V_k$. The
trigonometric R-matrix $\check{R}(x)\in{\rm End}(V\otimes V)$,
which satisfies the (graded) QYBE,
\beq
(I\otimes\check{R}(x))(\check{R}(xy)\otimes I)(I\otimes\check{R}(y))
 =(\check{R}(y)\otimes I)(I\otimes\check{R}(xy))(\check{R}(x)\otimes I),
\eeq
was given in \cite{Bra94a,Del94,Bra94b,others} in the form
\beq
\check{R}(x)=\frac{x-q^{2\a}}{1-xq^{2\a}}\check{P}_1+\check{P}_2+
  \frac{1-xq^{2\a+2}}{x-q^{2\a+2}}\check{P}_3.\label{trigono-R}
\eeq
Note, however, that $q$ and $\a$ are both free parameters which do not enter
the (graded) QYBE. Setting $x=q^\t$ and
taking the $q=1$ limit, one gets the corresponding rational R-matrix
(which also satisfies the (graded) QYBE)
\beq
\check{R}^{\rm r}(\t)=-\frac{\t-2\a}{\t+2\a}\check{P}_1^{(0)}+\check{P}_2^{(0)}
  -\frac{\t+2\a+2}{\t-2\a-2}\check{P}_3^{(0)}\label{rational-R}
\eeq
where $\check{P}^{(0)}_k,~k=1,2,3$, are classical ($q=1$) versions of
$\check{P}_k$, i.e. projection operator from $V^{(0)}\otimes V^{(0)}$
onto $V_k^{(0)}$, with $V^{(0)}$ and $V_k^{(0)}$ being the $q=1$ versions
of $V$ and $V_k$, respectively.  Note that $V^{(0)}$ and $V^{(0)}_k$
are actually $gl(2|1)$-modules, and $V^{(0)}$ affords the representation
(\ref{matrix}).  The projectors $\check{P}^{(0)}_k$
can easily be evaluated:
\bea
&&\check{P}_1^{(0)}=|\Psi_1^1\re\le\Psi_1^1|+|\Psi_2^1\re\le\Psi_2^1|
  +|\Psi_3^1\re\le\Psi_3^1|+|\Psi_4^1\re\le\Psi_4^1|,\no\\
&&\check{P}_3^{(0)}=|\Psi_1^3\re\le\Psi_1^3|+|\Psi_2^3\re\le\Psi_2^3|
  +|\Psi_3^3\re\le\Psi_3^3|+|\Psi_4^3\re\le\Psi_4^3|,\no\\
&&\check{P}_2^{(0)}=I-\check{P}_1^{(0)}-\check{P}_3^{(0)}\label{projectors}
\eea
where $|\Psi_k^1\re$ and $|\Psi_k^3\re,~k=1,2,3,4$,
form the symmetry adapted bases for the
spaces $V_1^{(0)}$ and $V_3^{(0)}$, respectively. Note that
$\check{R}^{\rm r}(0)\equiv I$.
We now compute $|\Psi_k^1\re$ and $|\Psi_k^3\re,~k=1,2,3,4$.
By means of the matrix representation (\ref{matrix}), one can show
\bea
&&|\Psi^1_1\re=|1\re\otimes |1\re,\no\\
&&|\Psi^1_2\re=\frac{1}{\sqrt{2}}(|2\re\otimes |1\re+|1\re\otimes |2\re),\no\\
&&|\Psi^1_3\re=\frac{1}{\sqrt{2}}(|3\re\otimes |1\re+|1\re\otimes |3\re),\no\\
&&|\Psi^1_4\re=\frac{1}{\sqrt{2(2\a+1)}}[\sqrt{\a+1}(|4\re\otimes |1\re
  +|1\re\otimes |4\re)+\sqrt{\a}(|2\re\otimes |3\re-|3\re\otimes |2\re)],\no\\
&&|\Psi^3_1\re=\frac{1}{\sqrt{2(2\a+1)}}[\sqrt{\a}(|4\re\otimes |1\re
  +|1\re\otimes |4\re)+\sqrt{\a+1}(-|2\re\otimes |3\re
  +|3\re\otimes |2\re)],\no\\
&&|\Psi^3_2\re=\frac{1}{\sqrt{2}}(|2\re\otimes |4\re+|4\re\otimes |2\re),\no\\
&&|\Psi^3_3\re=\frac{1}{\sqrt{2}}(|3\re\otimes |4\re+|4\re\otimes |3\re),\no\\
&&|\Psi^3_4\re=|4\re\otimes |4\re\label{basis}
\eea
which are easily seen to be orthonormal, so that
\bea
&&\le\Psi^1_k|=\left (|\Psi^1_k\re\right )^\dagger,~~~~
  \le\Psi^3_k|=\left (|\Psi^3_k\re\right )^\dagger,~~~k=1,2,3,4,\no\\
&&\left (|\b\re\otimes |\g\re\right )^\dagger=(-1)^{[|\b\re][|\g\re]}
  \left (|\b\re\right )^\dagger\otimes \left (|\g\re\right )^\dagger,\no\\
&&\left (|\b\re\right )^\dagger=\le\b|,~~~~~~\forall \b=1,2,3,4.\label{dual}
\eea
Here $[|\b\re]$ stands for the grading of the state $|\b\re$:~
$[|\b\re]=0$ for even (bosonic) $|\b\re$ and $[|\b\re]=1$ for odd
(fermionic) $|\b\re$. (Readers should keep in mind that the multiplication
rule for the tensor product is defined by
\beq
(a\otimes b)(c\otimes d)=(-1)^{[b][c]}\;(ac\otimes bd)
\eeq
for any elements $a,~b,~c$ and $d$.)

Using the rational R-matrix (\ref{rational-R}) and denoting
\beq
\check{R}^{\rm r}_{i,i+1}(\t)=I\otimes\cdots I\otimes
  \underbrace{\check{R}^{\rm r}(\t)}_{i~i+1}\otimes I
  \otimes\cdots\otimes I
\eeq
one may define \cite{Skl88} the local hamiltonian
\beq
H^{\rm R}_{i,i+1}(\a)=\left .\frac{d}{d\t}\check{R}^{\rm r}_{i,i+1}
   (\t)\right |_{\t=0}=
   -\frac{1}{\a}\left (\check{P}_1^{(0)}\right )_{i,i+1}+\frac{1}{\a+1}\left (
   \check{P}_3^{(0)}\right )_{i,i+1}.\label{h-r}
\eeq
By (\ref{projectors}), (\ref{basis}), (\ref{dual}) and (\ref{choice}), and
after tedious but straightforward manipulation, one gets, up to
a constant,
\beq
H^{\rm Q}_{i,i+1}(U)=-2(\a+1)H^{\rm R}_{i,i+1}(\a)\label{qld}
\eeq
which implies that the local hamiltonian $H^{\rm Q}_{i,i+1}(U)$ is
indeed derived from the $gl(2|1)$-invariant rational R-matrix which
satisfies the (graded) QYBE.
(Note that the identity (\ref{qld}) also indicates that $H^{\rm Q}(U)$
commutes with all the nine generators of $gl(2|1)$ since the rational
R-matrix $\check{R}^{\rm r}(\t)$ is a $gl(2|1)$ invariant.)

Now the exact solvability on the one-dimensional periodic lattice of our
model is seen as the following four steps. {\it Step 1}:
The hamiltonian $H^{\rm Q}(U)$ is self-adjoint and thus is diagonalizable.
{\it Step 2}: Relation (\ref{qld})
immediately makes it clear \cite{Skl88} that on the one dimensional
periodic lattice the global hamiltonian $H^{\rm Q}(U)$ commutes
with the transfer matrix $t(\t)$ constructed from
the rational R-matrix (\ref{rational-R})
(see c.f. \cite{Skl88} for the standard definition of the transfer matrix),
for any value of the parameter $\t$.
{\it Step 3}:
Using the fact, established in the {\it rational} case, that
$R^{\rm r}(\t)^\dagger=R^{\rm r}(\t)$, where
$R^{\rm r}(\t)=P\check{R}^{\rm r}(\t)$ and $P$ is the graded
permutation operator of the electron states (\ref{states})
, one may show that the
transfer matrix $t(\t)$ is self-adjoint and consequently diagonalizable
for any given parameter $\t$. (This result is in fact
established for real $\t$
but should also be valid for all complex $\t$ by using
analytical continuation arguments.) We remark here that the results in this
step are actually quite general: they are valid
for any (other) {\it rational} R-matrices arising from
{\it unitary} representations of any (other) quantum superalgebras.
{\it Step 4}: It can easily be shown that
$[t(\t), t(\t')]=0,~\forall \t,\t'$ and thus $t(\t)$ is diagonalizable
simultaneously for all $\t$.
Summarizing the above four steps, one sees that
the hamiltonian $H^{\rm Q}(U)$ satisfies the standard
requirement for a model to be exactly solvable by the Bethe ansatz. This
completes the proof for the exact solvability in one dimension of our model.
The details of solution of the model
is deferred to a separate publication.

\vskip.3in
We thank Tim Baker and Peter Jarvis for correspondences,
and V. Rittenberg for useful information.
We would like to express our sincere thanks to Gustav W. Delius, without
whose previous collaborations this work would not have been done.
We are grateful to F.H.L. Essler and V. Korepin for comments and suggestions.
The financial support from the Australian Research Council is gratefully
acknowledged.

\newpage

\end{document}